\documentstyle [12pt,epsf]{article}

\input{epsf}
\begin{document}
\begin{titlepage}
\begin{center}

 \vspace{-0.7in}

{\large \bf Stochastic Quantization \\ of \\ Real-Time Thermal Field Theory}\\
 \vspace{.3in}{\large\em T. C. de Aguiar\,\footnotemark[1], N.~F.~Svaiter \footnotemark[2]}\\
\vspace{.1in}
Centro Brasileiro de Pesquisas F\'{\i}sicas\, - CBPF,\\
 Rua Dr. Xavier Sigaud 150,\\
 Rio de Janeiro, RJ, 22290-180, Brazil \\

\vspace{.3in}{\large\em G. Menezes \,\footnotemark[3]}\\

\vspace{.1in}
Instituto de F\'{\i}sica Te\'{o}rica, Universidade Estadual Paulista,\\
 Rua Dr. Bento Teobaldo Ferraz 271, Bloco II, Barra Funda,\\
 S\~ao Paulo, SP,  01140-070, Brazil \\

\subsection*{\\Abstract}
\end{center}
\baselineskip .18in

We use the stochastic quantization method to obtain the free
scalar propagator of a finite temperature field theory formulated
in Minkowski spacetime. First we use the Markovian stochastic
quantization approach to present the two-point function of the
theory. Second, we assume a Langevin equation with a memory kernel
and Einstein's relations with colored noise. The convergence of
the stochastic processes in the asymptotic limit of the Markov
parameter of these Markovian and non-Markovian Langevin equations
for a free scalar theory is obtained. Our formalism can be the
starting point to discuss systems at finite temperature out of
equilibrium.

\vspace{0,3cm}
 PACS numbers: 03.70.+k, 11.10.Wx, 05.40.-a

\footnotetext[1]{e-mail:\,\,deaguiar@cbpf.br}
\footnotetext[2]{e-mail:\,\,nfuxsvai@cbpf.br}
\footnotetext[3]{e-mail:\,\,gsm@ift.unesp.br}

\end{titlepage}
\newpage\baselineskip .20
in
\section{Introduction}
\quad

The real time formalism allows one to discuss in detail finite
temperature field theory out of equilibrium. The aim of this paper
is to present an alternative approach to study finite temperature
field theory in the the real time formalism using the stochastic
quantization.

The program of stochastic quantization, first proposed by Parisi and
Wu \cite{parisi}, and the stochastic regularization were carried out
for systems described by fields defined in flat, Euclidean
manifolds. A brief introduction to stochastic quantization can be
found in Refs. \cite{sakita} \cite{ii} \cite{namiki1}, and a
complete review is given in Ref. \cite{damre}. Recently Menezes and
Svaiter \cite{menezes1} \cite{menezes3} implemented the stochastic
quantization to study systems with complex valued path integral
weights. Since we have a non-positive definite measure, the
convergence of the stochastic process in the asymptotic limit of the
Markov parameter is not achieved. To circumvent this problem, these
authors assumed a Langevin equation with memory kernel and
Einstein's relations with colored noise. In the asymptotic limit of
the Markov parameter, the equilibrium solution of such Langevin
equation was analyzed. It was shown that for a large class of
elliptic non-Hermitean operators, which define different models in
quantum field theory, the solution converges to the correct
equilibrium state in the asymptotic limit of the Markov parameter
$\tau\rightarrow\infty$.

We would like to remark that such kind of problems of non-convergence
of the Langevin equation in the stochastic quantization framework also
appears if someone consider the stochastic quantization of classical fields
defined in a generic curved manifold. For curved static manifolds,
the implementation of the stochastic quantization is straightforward. In
this situation it is possible to perform a Wick rotation, i.e., analytically extend the
pseudo-Riemannian manifold to the Riemannian domain without
problem. Recently, the stochastic quantization for fields defined
in a curved spacetime have been studied in the Refs.
\cite{jpa} \cite{cqg}. Nevertheless, for non-static curved
manifolds we have to extend the formalism beyond the Euclidean
signature, i.e., to formulate the stochastic quantization in
pseudo-Riemannian manifold, instead of formulating it in the Riemannian space, as was
originally proposed. See for example the discussion presented by
H\"{u}ffel and Rumpf \cite{huff} and Gozzi \cite{gozzi}. In the first
of these papers the authors proposed a modification of the
original Parisi-Wu scheme, introducing a complex drift term in the
Langevin equation, to implement the stochastic quantization in
Minkowski spacetime. Gozzi studied the spectrum of the
non-self-adjoint Fokker-Planck Hamiltonian to justify this
program. See also the papers \cite{huff2} \cite{call}. Of course,
these situations are special cases of ordinary Euclidean
formulation for systems with complex actions.

The main difference between the implementation of the stochastic
quantization in Minkowski spacetime and in Euclidean space is the
fact that in the latter case the approach to the equilibrium state
is a stationary solution of the Fokker-Planck equation. In the
Minkowski formulation, the Hamiltonian is non-Hermitian and the
eigenvalues of such Hamiltonian are in general complex. The real
part of such eigenvalues are important to the asymptotic behavior at
large Markov time, and the approach to the equilibrium is achieved
only if we can show its semi-positive definiteness. The crucial
question is the following: what happens if the Langevin equation
describes diffusion around a complex action? Some authors claim that
it is possible to obtain meaningful results out of Langevin equation
describing diffusion processes around a complex action. Parisi
\cite{con4} and Klauder and Peterson \cite{con5} investigated the
complex Langevin equation, where some numerical simulations in
one-dimensional systems were presented. See also the papers
\cite{con6} \cite{con7}. We would also like to mention the approach
developed by Okamoto et al. \cite{okamoto} where the role of the
kernel in the complex Langevin equation was studied. More recently,
Guralnik and Pehlevan constructed an effective potential for the
complex Langevin equation on a lattice \cite{gura}. These authors
also investigated a complex Langevin equation and Dyson-Schwinger
equations that appear in such situations \cite {sd}.

We would like to remind that there are many examples in the literature where
Euclidean action is complex. We
have, for example, $QCD$ with non-vanishing chemical potential at
finite temperature; for $SU(N)$ theories with $N>2$, the fermion
determinant becomes complex and also the effective action. Complex
terms can also appear in the Langevin equation for fermions, but a
suitable kernel can circumvent this problem \cite{f1} \cite{f2}
\cite{f3}. Another important case that deserves attention is the
stochastic quantization of topological field theories. The
simplest case though is, of course, the stochastic quantization in
Minkowski spacetime, as we discussed. This situation appears in
the case for non-equilibrium problems, which are not amenable to
an Euclidean formulation. Recently, Berges and Stamatescu \cite{berg} used
stochastic quantization techniques to present lattice simulations
of non-equilibrium quantum fields in Minkowskian spacetime.

In the perturbation theory in quantum field theory at finite
temperature there are three estabilished methods: the Matsubara
method \cite{mat1} \cite{mat2}, the path ordered method \cite{po1}
\cite{po2} and the Thermo Field Dynamics (TFD) approach
\cite{tfd1} \cite{tfd2}. For non-equilibrium quantum field
systems, the path ordered method or the Thermo Field Dynamics must
be used. In the Thermo Field Dynamics approach one can develop
finite temperature field theory in real time using the operator
formalism, while in the path ordered method the path integral
formalism is used. These three methods are related to each other
by the analytical continuation of time variables. The motivation
of this paper is to present an alternative approach to study
finite temperature field theory in the the real time formalism
using the Markovian and the non-Markovian stochastic quantization
procedures \cite{menezes2}. Basic ideas of the non-Markovian
Langevin equation can be found in the Refs. \cite{fox1}
\cite{fox2} \cite{z} \cite{kubo}.

The outline of the paper is the following. Introduction is given
in section I. In section II  we present a brief review of the real
time formalism in quantum field theory. In section III we use the
Markovian stochastic quantization method to study a
non-equilibrium thermal field theory formulated in Minkowski
spacetime. The non-Markovian approach of the stochastic
quantization applied to a thermal scalar field theory is developed
in section IV. Conclusions are given in section V. In the
appendix,  convergence conditions for the stochastic process are
derived. In this paper we use $\hbar=c=k_{B}=1$.

\section{Real-time formalism in finite temperature quantum field theory}

In this section we give a brief survey of the formulation of field
theory at finite temperature in Minkowski spacetime. Unlike in the
imaginary time formalism, in real time formulation, sums over
Matsubara frequencies are absent and there is no need to
analytically extend the Green functions back to Minkowski spacetime.
Moreover, the real time formalism is the starting point for the
development of the non-equilibrium quantum field theory, since the
investigation of dynamical properties of systems is more naturally
performed in this formalism. The real time formalism can describe
non-equilibrium processes because the time variable plays a
fundamental role and and cannot be traded in for an equilibrium
temperature.

For simplicity we work with a neutral scalar field. The field
operator in the Heisenberg picture is given by
\begin{equation}
\phi(t,\textbf{x})=e^{iHt}\phi(0,\textbf{x})\,e^{-iHt},
\label{eq1}
\end{equation}
where the time variable $t$ is allowed to be complex. The main
quantities to be computed are the thermal Green functions
$G_C(x_1,\ldots,x_N)$, defined as
\begin{equation}
G_C(x_1,\ldots,x_N)=\langle T_C\left(\phi(x_1)\ldots \phi(x_N)\right)\rangle_\beta, \label{eq2}
\end{equation}
where the time ordering is taken along a complex time path,
yet to be defined. Considering a parametrization $t=z(v)$ of the path,
the following expressions:
\begin{eqnarray}
\theta_C(t-t')&=&\theta(v-v'), \label{eq3} \\
\delta_C(t-t')&=&\left(\frac{\partial z}{\partial v}\right)^{-1}\delta(v-v'), \label{eq4}
\end{eqnarray}
define the generalized $\theta-$ and $\delta-$functions.
The functional differentiation is also extended in the following way:
\begin{equation}
\frac{\delta j(x)}{\delta j(x')}=\delta_C(t-t')\delta^3(\textbf{x}-\textbf{x}'), \label{eq5}
\end{equation}
for functions $j(x)$ defined on the path $C$. The Green functions
defined by Eq. (\ref{eq2}) can be obtained from a generating
functional $Z_C[\beta;j]$ through the expression
\begin{equation}
G_C(x_1,\ldots,x_N)=\frac{(-i)^{N}}{Z_C[\beta;j]} \frac{\delta^N
Z_C[\beta;j]}{\delta j(x_1)\cdots  \delta j(x_N)}\biggl|_{\,
j=0}. \label{eq6}
\end{equation}
in the above equation, the generating functional is given by
\begin{eqnarray}
&&Z[\beta;j]={\rm Tr}\left[e^{-\beta H}
\ T_C\exp{\left(i\int_C d^4x \ j(x)\phi(x)\right)}\right] \\ && \nonumber
=\int{\mathcal{D}}\phi' \ \langle\phi'(x);t-i\beta|
T_C \ \exp{\left(i\int_C d^4x \ j(x)\phi(x)\right)}|\phi'(x);t\rangle, \label{eq7}
\end{eqnarray}
where the path $C$ must go through all the arguments of the
Green functions. It is also possible to note from this expression
that the path $C$ starts from a time $t_i=t$ and ends at a time $t_f=t-i\beta$.
We may recast the generating functional into the form
\begin{equation}
Z_C[\beta;j]={\mathcal{N}}\exp{\left\{-i\int_C \ d^4x
\ V\left(\frac{\delta}{i\delta}j(x)\right)\right\}}Z_C^F[\beta;j] \label{eq8}
\end{equation}
where ${\mathcal{N}}$ is a normalization parameter and the free generating functional is given by
\begin{equation}
Z^F_C[\beta;j]= \exp{\left\{-\frac{1}{2}\int_C d^4x \ \int_C d^4y
\ j(x)D^F_C(x-y)j(y)\right\}}. \label{eq9}
\end{equation}
In Eq. (\ref{eq9}), the propagator $D^F_C(x-y)$ is defined through the
formula
\begin{equation}
D^F_C(x-x')=\theta_C(t-t')D^{>}_C(x,x')+\theta_C(t'-t)D^{<}_C(x,x'), \label{eq10}
\end{equation}
where $D^{>}_C(x,x')$ and $D^{<}_C(x,x')$ are, respectively:
\begin{eqnarray}
D^{>}_C(x,x')&=&\langle \phi(x)\phi(x')\rangle_\beta, \\ \nonumber
D^{<}_C(x,x')&=&\langle \phi(x')\phi(x)\rangle_\beta. \label{eq11}
\end{eqnarray}

Since the propagator $D^F_C(x-x')$ is properly defined in the
interval $-\beta\leq{\rm Im}(t-t')\leq \beta$, one may conclude
that the path considered must be such that the imaginary part of
the time variable $t$ is non-increasing when the parameter $v$
increases. Furthermore, since we are interested in Green functions
whose arguments are real, the path $C$ must contain the real axis.
One possible choice for the contour $C$ is described in the
following \cite{lebellac}:
\begin{enumerate}

\item $C$ starts from a real value $t_i$, large and negative.

\item The contour follows the real axis up to the large positive
value $-t_i$. This part of $C$ is denoted by $C_1$.

\item The path from $-t_i$ to $-t_i-i\frac{\beta}{2}$, along a
vertical straight line. This is denoted by $C_3$.

\item The path follows a horizontal line $C_2$  going from
$-t_i-i\frac{\beta}{2}$ to $t_i-i\frac{\beta}{2}$.

\item Finally, the path follows a vertical line $C_4$ from
$t_i-i\frac{\beta}{2}$ to $t_i-i\beta$.

\end{enumerate}

Taking $t_i\to -\infty$, the free generating functional can be factorized,
\begin{equation}
Z^F_C[\beta;j]=Z^F_{C_1\cup C_2}[\beta;j]Z^F_{C_3\cup C_4}[\beta;j]. \label{eq12}
\end{equation}
The Green functions with real time arguments can be deduced from
$Z^F_{C_1\cup C_2}[\beta;j]$ only. The $Z^F_{C_3\cup C_4}[\beta;j]$
generating functional can be considered a multiplicative constant.
Choosing $t$ and $t'$ real, running from $-\infty$ to $\infty$ and
label the sources $j_1(x)=j(t,\textbf{x})$ and $j_2(x)=j(t-i\beta/2,\textbf{x})$.
Also, one has $\displaystyle  \frac{\delta j_a(x)}{\delta j_b(x')}=\delta_{ab}\delta^4(x-x')$.
With this expressions one may rewrite the free generating functional as
\begin{equation}
Z_C^F[\beta;j]={\mathcal{N}}'\exp{\left\{-\frac{1}{2}
\int d^4x \ \int d^4x' \ j_a(x)D^F_{ab}j_{b}(x')\right\}}, \label{eq13}
\end{equation}
where, again, ${\mathcal{N}}'$ is a normalization parameter.
The components of the matricial propagator $D^{F}_{ab}(x-x')$ are
given by
\begin{eqnarray}
D_{11}^F(x-x')&=&D_F(t-t',\textbf{x}-\textbf{x}'), \\ \nonumber
D_{22}^F(x-x')&=&D_F^*(t-t',\textbf{x}-\textbf{x}'), \\ \nonumber
D_{12}^F(x-x')&=&D^{<}(t-t'+i\beta/2,\textbf{x}-\textbf{x}'), \\
\nonumber
D_{21}^F(x-x')&=&D^{>}(t-t'-i\beta/2,\textbf{x}-\textbf{x}').
\label{eq14}
\end{eqnarray}
The effective generating functional can be written as
\begin{eqnarray}
&&Z_C[\beta;j]=\int {\mathcal{D}}\phi_1{\mathcal{D}} \phi_2 \
\exp{\left\{-\frac{1}{2} \int d^4x \ d^4x' \
\phi_a(x)(D_F^{-1})_{ab}(x-x')\phi_b(x')\right\}} \\ && \nonumber
\times \exp{\left\{-i\int d^4x \ \left(V(\phi_1)-
V(\phi_2)\right)+i\int d^4x \ j_a(x)\phi_a(x)\right\}}. \label{eq15}
\end{eqnarray}
The field $\phi_2$ may be interpreted as a ghost field on the
contour $C_2$. This doubling of the field degrees of freedom,
which does not occur in imaginary time formulation, is unavoidable
in the real time formulation. For more details on this subject,
the reader is referred to the original paper of Niemi and Semenoff
\cite{niemi} or the Landsmann and van Weert review
{\cite{landsman}}.

\section{Real-time finite temperature quantum field theory: the Markovian stochastic quantization approach}

The real time formalism is a framework to describe both
equilibrium and non-equilibrium systems. Dynamical questions, as
for example a weakly interacting Bose gas having a temperature
gradient can be studied only in the real time formalism, with a
matrix structure of the propagator.  Before we study the
non-Markovian approach, in this section we will analyze the usual
stochastic quantization of a finite temperature field theory
formulated in Minkowski space. In Minkowski space, it is well
known that the Langevin equation should be written as:
\begin{equation}
\frac{\partial}{\partial \tau} \phi(x,\tau) = i\frac{\delta S}
{\delta \phi(x)}\biggl|_{\,\phi(x)=\phi(x,\tau)}+\eta(x,\tau),
\label{01}
\end{equation}
where $S(\phi)$ is the action for a free scalar field:
\begin{equation}
S(\phi)=\int d^dx \frac{1}{2}\left\{\partial^\mu\phi \
\partial_\mu\phi-m^2\phi^2\right\}, \label{02}
\end{equation}
and the correlation functions for the noise field are:
\begin{eqnarray}
\langle \eta(x,\tau) \rangle_\eta &=&0, \label{03} \\
\langle \eta(x,\tau)\eta(x',\tau') \rangle_\eta &=&
\delta(|\tau-\tau'|)\delta^d(x-x'). \label{04}
\end{eqnarray}
If we consider a complex free scalar field, with an action $S(\phi, \phi^{*})$ given
by
\begin{equation}
S=\int d^dx
\left\{\partial^\mu\phi^{*}\partial_\mu\phi-m^2\phi^{*}\phi\right\},
\label{05}
\end{equation}
we should have two Langevin equations, one for the scalar field
and the other to its complex conjugate. If we write $\Phi =
\pmatrix{\phi\cr \phi^{*}}$, and working in Fourier space, we can
write the Langevin equations as:
\begin{equation}
\frac{\partial}{\partial \tau} \Phi_a(k,\tau) = i
(D_{0}^{-1})_{ab}\Phi_{b}(k,\tau) + \eta_a(k,\tau), \label{06}
\end{equation}
where we also consider a complex noise field, $\eta =\pmatrix{\eta
\cr \eta^{*}}$, $a,b = 1, 2$ and
\begin{equation}
(D_{0}^{-1})_{ab}(k)= \pmatrix{\hfill (k^2 - m^2 + i\epsilon)& 0
\hfill\cr \hfill 0 & -(k^2 - m^2 - i\epsilon)\hfill\cr}. \label{07}
\end{equation}
The literature emphasizes \cite{damre} that the addition of a
negative imaginary mass term $-(i/2) \ \epsilon\phi^{*}\phi$ to
the action in Eq. (\ref{05}) is necessary in order to obtain
convergence for the stochastic process being considered. So, it
means that we can only take the limit $\epsilon\rightarrow 0$
after all calculations have been performed. That explains the
presence of the term $-i\epsilon$ in the expression for the
quantity $(D_{0}^{-1})_{ab}$, that appears in Eq. (\ref{07}). It
is straightforward to obtain the two-point correlation functions
for this case and to develop the perturbative solution to Eq.
(\ref{06}), with the following discussion on stochastic diagrams.
We do not wish to go into details here. For the interested reader,
we recommend the references \cite{huff} and \cite{huff2}.

Now, let us point our attentions to the non-equilibrium case. As
we discussed in the last section, the doubling of the field
degrees of freedom is unavoidable in the real time formulation.
So, we can also write the field as an isovector $\phi =
\pmatrix{\phi_1 \cr \phi_{2}}$. The action for this isovector
scalar field, in the free case, is given by:
\begin{equation}
S = \frac{1}{2} \int d^4x \
d^4x'\phi_a(x)(D_F^{-1})_{ab}(x-x')\phi_b(x'), \label{08}
\end{equation}
where the components of $(D_F)_{ab}$ are given by Eq. (14). In
Fourier space:
\begin{equation}
(D_F)_{ab}(k) = (U^{t})_{ac}(\theta)(D_{0})_{cd}(k)(U)_{db}(\theta),
\label{09}
\end{equation}
where $(D_{0})_{ab}$ is the inverse of $(D_{0}^{-1})_{ab}$, given by
Eq. (\ref{07}), and
\begin{equation}
(U)_{ab}(\theta)= \pmatrix{\hfill \cosh\theta & \sinh\theta
\hfill\cr \hfill \sinh\theta & \cosh\theta\hfill\cr}, \label{010}
\end{equation}
where:
\begin{equation}
\cosh^2\theta = \frac{e^{\beta |k_0|}}{e^{\beta |k_0|} - 1}.
\label{011}
\end{equation}
So, we can split $(D_F)_{ab}$ into two parts, $(D_F)_{ab} = (D_0)_{ab} +
(D_{\beta})_{ab}$, where $(D_0)_{ab}(k)$ is temperature independent
\begin{equation}
(D_{0})_{ab}(k)= \pmatrix{\hfill \frac{1}{k^2 - m^2 + i\epsilon}&
0 \hfill\cr \hfill 0 & \frac{-1}{k^2 - m^2 - i\epsilon}\hfill\cr},
\label{012}
\end{equation}
and all temperature dependence appears in $(D_{\beta})_{ab}(k)$, which is given by
\begin{equation}
(D_{\beta})_{ab}(k)= \frac{-i\epsilon}{(k^2 - m^2)^2 +
\epsilon^2}\pmatrix{\hfill 2\sinh^2\theta & \sinh2\theta \hfill\cr
\hfill \sinh2\theta & 2\sinh^2\theta\hfill\cr}. \label{013}
\end{equation}
Notice that, in the limit $\epsilon\rightarrow 0$, we have:
\begin{equation}
\frac{\epsilon}{(k^2 - m^2)^2 + \epsilon^2}\rightarrow \pi\delta(k^2
- m^2). \label{014}
\end{equation}
As in the zero temperature case, we have to use the expressions
above with finite $\epsilon$ and take the $\epsilon\rightarrow 0$
limit after all the calculations have been done in order to obtain
convergence in the limit $\tau\rightarrow\infty$. We also know that
this is the case for the path integral formalism \cite{niemi}.
The Markovian Langevin equation for the non-equilibrium case is
given by
\begin{equation}
\frac{\partial}{\partial \tau} \phi_a(k,\tau) = i
(D_{F}^{-1})_{ab}(k)\phi_{b}(k,\tau) + \eta_a(k,\tau), \label{015}
\end{equation}
where the noise correlation functions are given by:
\begin{eqnarray}
\langle \eta_a(k,\tau) \rangle_\eta &=&0, \label{03} \\
\langle \eta_a(k,\tau)\eta_b(k',\tau') \rangle_\eta &=&
(2\pi)^d\,\delta_{ab}\,\delta^d(k+k')\delta(|\tau-\tau'|).
\label{016}
\end{eqnarray}
The solution for the Eq. (\ref{015}) is given by:
\begin{equation}
\phi_a(k,\tau) = \int^{\infty} d\tau'(g(k,\tau-\tau'))_{ab} \
\eta_b(k,\tau'), \label{017}
\end{equation}
where $g(k,\tau) = e^{iD_{F}^{-1}(k)\tau}\theta(\tau)$ is the
Green function for the diffusion problem. In order to check
convergence, i.e, to analyze if
$g(k,\tau)|_{\tau\rightarrow\infty}\rightarrow 0$, we must first
diagonalize the matrix $iD_{F}^{-1}(k)$. From Eqs. (\ref{012}) and
(\ref{013}), we have:
\begin{equation}
D_{F}^{-1}(k) = I(k,\epsilon)\pmatrix{\hfill \frac{-1}{k^2 - m^2 -
i\epsilon} - \frac{i\epsilon}{(k^2 - m^2)^2 +
\epsilon^2}\,2\,\sinh^2\theta & \frac{i\epsilon}{(k^2 - m^2)^2 +
\epsilon^2}\,\sinh2\theta \hfill\cr \hfill \frac{i\epsilon}{(k^2 -
m^2)^2 + \epsilon^2}\,\sinh2\theta & \frac{1}{k^2 - m^2 +
i\epsilon} - \frac{i\epsilon}{(k^2 - m^2)^2 +
\epsilon^2}\,2\,\sinh^2\theta\hfill\cr},
\end{equation}
where:
\begin{equation}
I(k,\epsilon)= \frac{-(k^2 - m^2)^2 - \epsilon^2(\cosh^4\theta +
\sinh^4\theta) - \frac{\epsilon^2}{2}\,\sinh^{2}2\theta}{((k^2 -
m^2)^2 + \epsilon^2)}.
\end{equation}
Diagonalizing $iD_{F}^{-1}(k)$, we get the matrix $D'(k)$, given
by:
\begin{equation}
D'(k) = i\,I(k,\epsilon)\pmatrix{\hfill \lambda_{+} & 0 \hfill\cr
\hfill 0 & \lambda_{-}\hfill\cr}, \label{a}
\end{equation}
where:
\begin{equation}
\lambda_{\pm} = \frac{\pm\sqrt{(k^2 - m^2)^2 -
\epsilon^2\,\sinh^{2}2\theta} - i\,\epsilon(1 +
2\sinh^2\theta)}{(k^2 - m^2)^2 + \epsilon^2}. \label{b}
\end{equation}
Since $I(k,\epsilon) < 0$, we notice from the above equations
that, indeed, we get
$g(k,\tau)|_{\tau\rightarrow\infty}\rightarrow 0$. We also remark
that, as in the zero temperature case, the convergence of the
stochastic process was possible because we have maintained in the
Eqs. (\ref{012}) and (\ref{013}) a finite $\epsilon$. As the
reader can easily verify from the Eqs. (\ref{a}) and (\ref{b}), if
we take the $\epsilon\rightarrow 0$ limit in the beginning of the
calculations, we should lose the convergence factor
$e^{-I(k,\epsilon)\epsilon(1 + 2\sinh^2\theta)}$. So this limit
should be taken after all the calculations have been done in order
to obtain convergence in the limit $\tau\rightarrow\infty$. We
also know that this is the case for the path integral formalism
\cite{niemi}.

Now, we are ready to calculate the two point function $\langle
\phi_a(k,\tau)\phi_b(k',\tau) \rangle_\eta$. Proceeding with
similar calculations as the zero temperature case, it is possible
to show that the two-point correlation function is given by:
\begin{equation}
\langle \phi_a(k,\tau)\phi_b(k',\tau) \rangle_\eta = (2\pi)^d
\delta^d(k + k')i\,(D_{F})_{ac}(k)(1 -
e^{2iD_{F}^{-1}(k)\tau})_{cb}, \label{018}
\end{equation}
so we see that, in the limit $\tau\rightarrow\infty$, we recover
the usual result. We are interested now to see the effects of a
memory kernel in this non-equilibrium quantum field theory. This
is the subject of the next section.

\section{Real-time finite temperature quantum field theory:
the non-Markovian stochastic quantization approach}

The aim of this section is to study finite temperature quantum
field theory in Minkowski spacetime, using the non-Markovian
stochastic quantization approach. In Minkowski space, the Langevin
equation with memory kernel is written as
\begin{equation}
\frac{\partial}{\partial \tau} \phi(x,\tau) = i\int_{0}^{\tau}ds \
M_\Lambda(\tau-s)\frac{\delta S} {\delta
\phi(x)}\biggl|_{\,\phi(x)=\phi(x,s)}+ \ \eta(x,\tau), \label{1}
\end{equation}
where $S$ is the action for the free scalar field, given by Eq.
(\ref{08}). The noise field distribution is such that its first
and second momenta are given by
\begin{eqnarray}
\langle \eta_a(x,\tau) \rangle_\eta &=&0, \label{2} \\
\langle \eta_a(x,\tau)\eta_b(x',\tau') \rangle_\eta &=&
2\delta_{ab}\,M_\Lambda(|\tau-\tau'|)\delta^d(x-x'), \label{3}
\end{eqnarray}
that is, the distribution is a colored noise Gaussian distribution.
We remind the reader that Eq. (\ref{1}) is to be understood as a
matrix equation.

Using a Fourier decomposition for the scalar and noise fields, given
by
\begin{equation}
X(x,\tau)=\frac{1}{(2\pi)^{\frac{d}{2}}}\int d^dk \ e^{ikx} X(k,\tau), \label{5}
\end{equation}
where the field $X$ represents either the noise field $\eta$ and the scalar field $\phi$,
we obtain that each Fourier mode $\phi(k,\tau)$ satisfies a Langevin equation of the form
\begin{equation}
\frac{\partial}{\partial \tau}\phi_a(k,\tau)=i\int_0^\tau ds \
M_\Lambda(\tau-s)\,(D_{F}^{-1})_{ab}(k)\,\phi_b(k,s) +
\eta_a(k,\tau), \label{6}
\end{equation}
where $D_{F}^{-1}(k)$ is the inverse of $D_{F}(k)$, defined by Eq.
(\ref{09}). With this decomposition, we obtain from Eq. (\ref{2})
and Eq. (\ref{3}) the following relations for the noise field
Fourier components,
\begin{eqnarray}
\langle \eta_a(k,\tau) \rangle_\eta &=&0, \label{7} \\
\langle \eta_a(k,\tau)\eta_b(k',\tau') \rangle_\eta &=&
2(2\pi)^d\,\delta_{ab}\, M_\Lambda(|\tau-\tau'|)\delta^d(k+k').
\label{8}
\end{eqnarray}
Defining the Laplace transform of the memory kernel as
\begin{equation}
M(z)=\int_0^\infty d\tau \ M_{\Lambda}(\tau)e^{-z\tau}, \label{9}
\end{equation}
we obtain the solution for Eq. (\ref{6}), subject to the initial
condition $\phi_a(k,\tau=0)=0$, $a = 1,2$:
\begin{equation}
\phi_a(k,\tau)=\int_0^\infty d\tau' \
G_{ab}(k,\tau-\tau')\eta_a(k,\tau') \label{10}
\end{equation}
In Eq. (\ref{10}) $G_{ab}(k,\tau-\tau') =
\Omega_{ab}(k,\tau-\tau')\theta(\tau-\tau')$ is the retarded Green
function for the diffusion problem and each component of the
$\Omega$-matrix is defined through its Laplace transform,
\begin{equation}
\Omega_{ab} = \biggl[z\delta_{ab} -
i\,M(z)\,\biggl(D_{F}^{-1}\biggr)_{ab}(k) \biggr]^{-1}. \label{11}
\end{equation}
Thus, the two point correlation function in the Fourier
representation is written as
\begin{eqnarray}
&&\langle \phi_a(k,\tau)\phi_b(k',\tau') \rangle_\eta =
D_{ab}(k,\tau,\tau') \nonumber \\&&
=2(2\pi)^d\delta^d(k+k')\int_0^{\tau} ds \int_0^{\tau'} ds' \
\left[\Omega(k,\tau-\tau')
\Omega(k,\tau-s')\right]_{ab}\,M_\Lambda(|s-s'|). \label{12}
\end{eqnarray}
The two-dimensional Laplace transform of the above equation is given by
\begin{eqnarray}
&&\int_0^\infty d\tau \ e^{-z\tau} \ \int_0^\infty d\tau' \
e^{-z'\tau'} \ \int_0^{\tau} ds \ \int_0^{\tau'} ds' \
\left[\Omega(k,\tau-\tau')\Omega(k,\tau-s')\right]_{ab}
\,M_\Lambda(|s-s'|) \nonumber
\\&& = \left[\Omega(k,z)\Omega(k',z')\right]_{ab}\left(\frac{M(z)+M(z')}{z+z'}\right).
\label{15}
\end{eqnarray}
Using the Eq. (\ref{11}) this expression becomes
\begin{eqnarray}
&&\int_0^\infty d\tau \ e^{-z\tau} \ \int_0^\infty d\tau' \
e^{-z'\tau'} \ \int_0^{\tau} ds \ \int_0^{\tau'} ds' \
\left[\Omega(k,\tau-\tau')
\Omega(k,\tau-s')\right]_{ab}\,M_\Lambda(|s-s'|) \nonumber
\\&& =
i\left(\frac{\Omega(k,z)+\Omega(k,z')}{z+z'}-\Omega(k,z)\Omega(k,z')\right)_{ac}(D_F)_{cb}(k).
\label{16}
\end{eqnarray}
Applying the inverse transform, we obtain for the two-point function
\begin{equation}
D_{ab}(k,\tau,\tau')=2i(2\pi)^d\,\delta^d(k+k')
\left(\Omega(k,|\tau' - \tau|) -
\Omega(k,\tau)\Omega(k,\tau')\right)_{ac}(D_F)_{cb}(k). \label{17}
\end{equation}

In order to investigate the convergence of the above equation, we need to
specify an expression for the memory kernel $M_\Lambda$. We set
\begin{equation}
M_\Lambda(\tau)=\frac{1}{2}\Lambda^2 e^{-\Lambda^2|\tau|}. \label{18}
\end{equation}

Substituting the Laplace transform of the Eq. (\ref{18}) in the
Eq. (\ref{11}), we have that the $\Omega$-matrix is given by
\begin{equation}
\Omega(k,\tau)=\pmatrix{\hfill \Omega_{11}(k,\tau) &
\Omega_{12}(k,\tau) \hfill\cr \hfill \Omega_{21}(k,\tau) &
\Omega_{22}(k,\tau) \hfill\cr}, \label{19}
\end{equation}
where the components $\Omega_{ab}(k,\tau)$ are given in the
Appendix. So, we are in a position to present an expression for the
two-point correlation function in the limit $\tau =
\tau'\rightarrow\infty$:
\begin{equation}
D_{ab}(k,\tau,\tau')|_{\tau =
\tau'\rightarrow\infty}=\,i\,\,(2\pi)^d\delta^d(k+k')(D_F)_{ab}(k),
\label{21}
\end{equation}
so that, in the limit $\epsilon\rightarrow 0$, we have:
\begin{equation}
D_{ab}(k,\tau,\tau')|_{\tau = \tau'\rightarrow\infty;
\,\epsilon\rightarrow 0}=\,i\,
\,(2\pi)^d\delta^d(k+k')(D_F)_{ab}(k)|_{\,\epsilon\rightarrow 0}.
\label{22}
\end{equation}

A question still remains opened. What are, if any, the advantages of
our non-Markovian method over the usual Markovian one? In order to
answer such question, we shall apply a Fokker-Planck analysis. As we
know, correlation functions are introduced as averages over $\eta$:
\begin{eqnarray}
&&\langle\,\phi
(x_{1},\tau_{1})\phi(x_{2},\tau_{2})\cdots\phi(x_{n},\tau_{n})\,\rangle_{\eta}=\nonumber\\
&& \emph{N}\int\,[d\eta]\exp\biggl(-\frac{1}{4} \int d^{d}x \int
d\tau\,\eta^{2}(x,\tau)\bigg)\,\phi(x_{1},\tau_{1})
\phi(x_{2},\tau_{2})\cdots\phi(x_{n},\tau_{n}), \label{s12}
\end{eqnarray}
where $\phi$ obeys Eq. (\ref{01}) and $\emph{N}$ is given by:
\begin{equation}
\emph{N}^{-1} = \int\,[d\eta]\exp\biggl(-\frac{1}{4} \int d^{d}x \int
d\tau\,\eta^{2}(x,\tau)\bigg)
\end{equation}

An alternative way to write this average is to introduce the
probability density $P[\phi,\tau]$, which is defined as
\cite{ili}:
\begin{equation}
P[\phi,\tau]\equiv\int\,[d\eta]\exp\biggl(-\frac{1}{4} \int
d^{d}x \int
d\tau\,\eta^{2}(x,\tau)\bigg)\,\prod\limits_{y}\delta(\phi(y)-\phi(y,\tau)).\label{s13}
\end{equation}
In terms of $P$, the correlation functions will read:
\begin{equation}
\langle\,\phi
(x_{1},\tau_{1})\phi(x_{2},\tau_{2})\cdots\phi(x_{n},\tau_{n})\,\rangle_{\eta}=
\emph{N}\int\,[d\phi]\,\phi(x_{1},\tau_{1})
\phi(x_{2},\tau_{2})\cdots\phi(x_{n},\tau_{n})P[\phi,\tau].
\label{s14}
\end{equation}
The free probability density $P$ satisfies the following Fokker-Planck equation:
\begin{equation}
\frac{\partial}{\partial\tau}P[\phi,\tau]=\int\,d^{d}x
\frac{\delta}{\delta\,\phi(x)}\Biggl(\frac{\delta}{\delta\,\phi(x)}-
i\frac{\delta\,S}{\delta\,\varphi(x)}\Biggr)P[\phi,\tau],
\label{s15}
\end{equation}
where $S$ is given by Eq. (\ref{02}) and with the initial condition:
\begin{equation}
P[\phi,0] = \prod\limits_{y}\delta(\phi(y)).
\end{equation}
The stochastic quantization says that we shall have:
\begin{equation}
w.\lim\limits_{\tau\rightarrow\infty}P[\phi,\tau] =
\frac{\exp\bigl(i\,S[\phi]\bigr)}
{\int\,[d\phi]\exp\bigl(i\,S[\phi]\bigr)},
 \label{s16}
\end{equation}
where the limit is supposed to be taken "weakly" in the sense of the
reference \cite{ili}.

In our real time non-Markovian case, if we notice the resemblance
between our retarded Green function $G_{ab}(k,\tau)$ and the one
found in Ref. \cite{menezes2}, we may follow similar steps to
calculate the free probability density. It is given by, in momentum
space:
\begin{equation}
P[\phi,\tau]=
N^{-1}\exp{\biggl(\frac{i}{2}\int\,dk\,\phi_a(k)D_{ab}^{-1}(k,\tau,\tau)\phi_b(-k)\biggr)}
\end{equation}
where $N^{-1}$ is a normalization factor and
$D_{ab}^{-1}(k,\tau,\tau')$ is the inverse of
$D_{ab}(k,\tau,\tau')$, defined by Eq. (\ref{17}). It is easy to
verify that, in the limit $\tau\rightarrow\infty$, $P[\phi,\tau]$
will satisfy, up to constants, similar relations as obtained by Ref.
\cite{ili}. However, for massless scalar theories, those estimations
in such reference decay as inverse power of $\tau$. In our approach,
even in the massless situation, an exponential behavior is found.
Therefore, in the limit $\tau\rightarrow\infty$, it seems that we
get an improved convergence.

In conclusion, we have used the stochastic quantization method to
study thermal field theory formulated in Minkowski spacetime. We
have assumed a Langevin equation with a memory kernel and
Einstein's relations with colored noise. From the above last
equation, we see that the equilibrium solution in the asymptotic
Markov time of this non-Markovian Langevin equation for scalar
theories can be obtained. Our approach based in stochastic
quantization using a non-Markovian Langevin equation proved to be
well suited to quantize a classical field out of equilibrium in
the real time formalism at finite temperature.

\section{Conclusions}
\quad

In the past several years there have been a lot of interest in
quantum field theory at finite temperature. There are three main
formulations of finite temperature field theory. The imaginary
time approach or the Matsubara formalism and the real time
formalism, which can be operatorial or use the path integral
approach. In the real time formalism, it is necessary to double
the number of the field degrees of freedom. To quantize a
classical thermal field theory out of equilibrium, using the
stochastic quantization, we are forced to work in the Minkowski
spacetime, where naturally a imaginary drift term appears in the
Langevin equation. Since in this case the path integral weight is
not positive definite, the stochastic quantization in this
situation is problematic. Parisi and Klauder proposed complex
Langevin equations \cite{con4} \cite{con5}, and some problems of
this approach are the following. First of all, complex Langevin
simulations do not converge to a stationary distribution in many
situations. Besides, if it does, it may converge to many different
stationary distributions. The complex Langevin equation also
appears when the original method proposed by Parisi and Wu is
extended to include theories with fermions \cite{f1} \cite{f2}
\cite{f3}. The first question that appears in this context is if
make sense the Brownian problem with anticommutating numbers. It
can be shown that, for massless fermionic fields, there will not
be a convergence factor after integrating the Markovian Langevin
equation. Therefore the equilibrium is not reached. One way of
avoiding this problem is to introduce a kernel in the Langevin
equation describing the evolution of two Grassmannian fields.

In this paper, we have used the method of the stochastic
quantization to study thermal field theory formulated in real
time. As we discussed, this closed time path method can be used to
describe non-equilibrium thermal field theory. First we use the
Markovian stochastic quantization approach to present the
two-point function of the theory. Second, we assumed a Langevin
equation with a memory kernel and Einstein's relation with colored
noise. The equilibrium solution of such Langevin equation was
analyzed. We have shown that for a large class of elliptic
non-Hermitean operators which define different models in quantum
field theory converges in the asymptotic limit of the Markov
parameter $\tau\rightarrow\infty$, and we have obtained the free
Green functions of the theory. Although non-trivial, the method
proposed can be extended to interacting field theory with complex
actions, where a consistent perturbation theory out of equilibrium
can be developed.

Finally, the literature has emphasized that the stochastic
quantization is only an alternative formalism to quantize a
classical field theory, but new results have not been obtained.
Nevertheless the stochastic quantization and the Langevin equation
can be extremely useful in numerical simulations of field theory
models \cite{lattice} \cite{lattice2}. The implementation of this
non-Markovian Langevin equation on the lattice is under
investigation by the authors.

\section{Acknowlegements}

We would like to thank prof. Gast\~ao Krein for many helpful
discussions. N. F. Svaiter would like to acknowledge the hospitality
of the Instituto de F\'{\i}sica Te\'{o}rica, Universidade Estadual
Paulista, where part of this paper was carried out. This paper was
supported by Conselho Nacional de Desenvolvimento Cientifico e
Tecnol{\'o}gico do Brazil (CNPq) and Funda\c{c}\~ao de Amparo a
Pesquisa de S\~ao Paulo (FAPESP).

\begin{appendix}
\makeatletter \@addtoreset{equation}{section} \makeatother
\renewcommand{\theequation}{\thesection.\arabic{equation}}

\section{Appendix}
In this appendix, we derive the components $\Omega_{ab}(k,\tau)$.
We can write the $\Omega$-matrix as an inverse of the matrix $A$,
whose components are given by
\begin{equation}
A = \pmatrix{\hfill z - i\,M(z)\,d\,' &  i\,M(z)\,b\,' \hfill\cr
\hfill i\,M(z)\,b\,' & z - i\,M(z)\,a\,' \hfill\cr}. \label{ap1}
\end{equation}
The quantities that appear in the $A$-matrix are defined by
\begin{eqnarray}
a\,' = \frac{a}{ad - b^2}, \label{ap2} \\
b\,' = \frac{b}{ad - b^2}, \label{ap3} \\
d\,' = \frac{d}{ad - b^2}, \label{ap4}
\end{eqnarray}
and
\begin{eqnarray}
a &=& \frac{1}{k^2 - m^2 + i\epsilon} -
\frac{i\epsilon}{(k^2 - m^2)^2 + \epsilon^2}\,2\,\sinh^2\theta, \label{ap5} \\
b &=& \frac{-i\epsilon}{(k^2 - m^2)^2 + \epsilon^2}\,\sinh2\theta, \label{ap6} \\
d &=& \frac{-1}{k^2 - m^2 - i\epsilon} - \frac{i\epsilon}{(k^2 -
m^2)^2 + \epsilon^2}\,2\,\sinh^2\theta. \label{ap7}
\end{eqnarray}
So, we will have
\begin{equation}
(\Omega)_{ab}(k,z) = (A^{-1})_{ab} = \pmatrix{\hfill
\Omega_{11}(k,z) & \Omega_{12}(k,z) \hfill\cr \hfill
\Omega_{21}(k,z) & \Omega_{22}(k,z) \hfill\cr}, \label{ap8}
\end{equation}
where:
\begin{eqnarray}
\Omega_{11}(k,z) &=& \frac{z - i\,M(z)\,a\,'}{(z - i\,M(z)\,a\,')(z
-
i\,M(z)\,d\,') + M^{2}(z)b\,'^{2}}, \label{ap9} \\
\Omega_{12}(k,z) = \Omega_{21}(z) &=& \frac{-i\,M(z)\,b\,'}{(z -
i\,M(z)\,a\,')(z -
i\,M(z)\,d\,') + M^{2}(z)b\,'^{2}}, \label{ap10} \\
\Omega_{22}(k,z)  &=& \frac{z - i\,M(z)\,d\,'}{(z - i\,M(z)\,a\,')(z
- i\,M(z)\,d\,') + M^{2}(z)b\,'^{2}}. \label{ap11}
\end{eqnarray}
The Laplace transform for the memory kernel, Eq. (\ref{18}), is
given by
\begin{equation}
M(z) = \frac{\,\,\Lambda^2}{2}\frac{1}{z + \Lambda^2}.
\end{equation}
So, inserting this result in Eqs. (\ref{ap9}), (\ref{ap10}) and
(\ref{ap11}), we get:
\begin{eqnarray}
\Omega_{11}(k,z) &=& \frac{P(z,t_a)}{Q(z)}, \label{ap12} \\
\Omega_{12}(k,z) = \Omega_{21}(z) &=& \frac{-(t_b\,z + t_b\,\Lambda^2)}{Q(z)}, \label{ap13} \\
\Omega_{22}(k,z)  &=& \frac{P(z,t_d)}{Q(z)}, \label{ap14}
\end{eqnarray}
where $t_j = i\,\frac{j\,'\,\Lambda^2}{2}$, $j = a, b, d$, and
\begin{eqnarray}
P(z,t_j) &=& z^3 + 2\,\Lambda^2\,z^2 + (\Lambda^4 - t_j)\,z - t_j\,\Lambda^2, \label{ap15} \\
Q(z)  &=& z^4 + 2\,\Lambda^2\,z^3 + (\Lambda^4 - u)z^2 -
u\,\Lambda^2\,z + v, \label{ap16}
\end{eqnarray}
with $u = \frac{i(a\,' + \, d\,')\Lambda^2}{2}$ and $v =
\frac{(\,b\,'^2 - \,a\,'d\,'\,)\Lambda^4}{4}$. From Eqs.
(\ref{ap2}), (\ref{ap3}) and (\ref{ap4}), we have that:
\begin{equation}
u = \frac{\Lambda^2\,\epsilon(1 + 2\sinh^2\theta)((k^2 - m^2)^2 +
\epsilon^2)}{-(k^2 - m^2)^2 - \epsilon^2(\cosh^4\theta +
\sinh^4\theta) + \frac{\epsilon^2}{2}\,\sinh^{2}2\theta},
\label{ap16a}
\end{equation}
and
\begin{equation}
v = -\frac{\Lambda^4}{4}\frac{((k^2 - m^2)^2 + \epsilon^2)}{-(k^2 -
m^2)^2 - \epsilon^2(\cosh^4\theta + \sinh^4\theta) +
\frac{\epsilon^2}{2}\,\sinh^{2}2\theta},
\end{equation}
so $u < 0$ and $v > 0$. In order to get the inverse Laplace
transform of each component of the $\Omega$-matrix, we must seek
for the solutions of the quartic equation $Q(z)=0$. As it is well
known, a general quartic equation is a fourth-order polynomial
equation of the form
\begin{equation}
z^4 + a_3 z^3 + a_2 z^2 + a_1 z + a_0 = 0. \label{ap17}
\end{equation}
Using the familiar algebraic technique \cite{abram}, it is easy to
show that the roots of Eq.(\ref{ap17}) are given by:
\begin{equation}
z_1 = -\frac{1}{4} a_3 + \frac{1}{2} R + \frac{1}{2} D, \label{ap18}
\end{equation}
\begin{equation}
z_2 = -\frac{1}{4} a_3 + \frac{1}{2} R - \frac{1}{2} D, \label{ap19}
\end{equation}
\begin{equation}
z_3 = -\frac{1}{4} a_3 - \frac{1}{2} R + \frac{1}{2} E, \label{ap20}
\end{equation}
\begin{equation}
z_4 = -\frac{1}{4} a_3 - \frac{1}{2} R - \frac{1}{2} E, \label{ap21}
\end{equation}
where:
\begin{equation}
R\equiv\biggl(\frac{1}{4} a_{3}^2 - a_2 + y_1\biggr)^{1/2},
\label{ap22}
\end{equation}
\begin{equation}
D\equiv \left\{ \biggl(F(R)+G\biggr)^{1/2} \hfill\hbox{for $R\neq0$}
\atop \biggl(F(0)+H\biggr)^{1/2} \quad \hbox{for $R=0,$}\right.
\label{ap23}
\end{equation}
\begin{equation}
E\equiv \left\{ \biggl(F(R)-G\biggr)^{1/2} \hfill\hbox{for $R\neq0$}
\atop \biggl(F(0)-H\biggr)^{1/2} \quad \hbox{for $R=0,$}\right.
\label{ap24}
\end{equation}
\begin{equation}
F(R)\equiv\frac{3}{4} a_{3}^2 - R^2 - 2a_2, \label{ap25}
\end{equation}
\begin{equation}
H\equiv2\biggl(y_{1}^2-4a_0\biggr)^{1/2}, \label{ap26}
\end{equation}
\begin{equation}
G\equiv\frac{1}{4}(4 a_3a_2-8a_1-a_{3}^3)R^{-1}, \label{ap27}
\end{equation}
and $y_1$ is a real root of the following cubic equation:
\begin{equation}
y^3 - a_2 y^2 + (a_1a_3 - 4a_0)y + (4a_2a_0 - a_{1}^2 -
a_{3}^2a_0)=0. \label{ap28}
\end{equation}
For convenience, let us assume that $R$, defined by Eq.(\ref{ap22}),
does not vanish. Comparing Eqs. (\ref{ap16}) and (\ref{ap17}), we
easily see that $a_3 = 2\,\Lambda^2$, $a_2= \Lambda^4 - u$, $a_1 =
-u\,\Lambda^2$ and $a_0 = v$. Therefore, the inverse Laplace
transform of $\Omega_{ab}$ is given by:
\begin{eqnarray}
\Omega_{11}(k,\tau) &=& \frac{P(z_1,t_a)}{(z_1 -
z_2)(z_1 - z_3)(z_1 - z_4)} e^{z_1 \tau} + \nonumber\\
 &&+\frac{P(z_2,t_a)}{(z_2 - z_1)(z_2 - z_3)(z_2 - z_4)}
e^{z_2 \tau} + \nonumber\\
 &&+\frac{P(z_3,t_a)}{(z_3 - z_1)(z_3 - z_2)(z_3 - z_4)}
e^{z_3 \tau} + \nonumber\\
&&+\frac{P(z_4,t_a)}{(z_4 - z_1)(z_4 - z_2)(z_4 - z_3)} e^{z_4 \tau}
, \label{ap29}
\end{eqnarray}
\begin{eqnarray}
\Omega_{12}(k,\tau) = \Omega_{21}(k,\tau) &=& -\Biggl(\frac{t_b\,z_1
+ t_b\,\Lambda^2}{(z_1 -
z_2)(z_1 - z_3)(z_1 - z_4)} e^{z_1 \tau} + \nonumber\\
 &&+\frac{t_b\,z_2 + t_b\,\Lambda^2}{(z_2 - z_1)(z_2 - z_3)(z_2 - z_4)}
e^{z_2 \tau} + \nonumber\\
 &&+\frac{t_b\,z_3 + t_b\,\Lambda^2}{(z_3 - z_1)(z_3 - z_2)(z_3 - z_4)}
e^{z_3 \tau} + \nonumber\\
&&+\frac{t_b\,z_4 + t_b\,\Lambda^2}{(z_4 - z_1)(z_4 - z_2)(z_4 -
z_3)} e^{z_4 \tau}\Biggr) , \label{ap30}
\end{eqnarray}
and, finally, $\Omega_{22}(k,\tau) = \Omega_{11}(k,\tau;
t_a\rightarrow t_d)$. The roots $z_i$ are given by:
\begin{equation}
z_1 = -\frac{\Lambda^2}{2} + \frac{1}{2}i \sigma + \frac{1}{2}
i\gamma, \label{ap31}
\end{equation}
\begin{equation}
z_2 = -\frac{\Lambda^2}{2} + \frac{1}{2} i\sigma - \frac{1}{2}
i\gamma, \label{ap32}
\end{equation}
\begin{equation}
z_3 = -\frac{\Lambda^2}{2} - \frac{1}{2} i\sigma + \frac{1}{2}
i\gamma, \label{ap33}
\end{equation}
\begin{equation}
z_4 = -\frac{\Lambda^2}{2} - \frac{1}{2} i\sigma- \frac{1}{2}
i\gamma, \label{ap34}
\end{equation}
with $\sigma = (|u| - \,y_1)^{1/2}$ and $\gamma = (-\Lambda^4 + |u|
+ y_1)^{1/2}$ being real quantities. The Eqs. (\ref{ap29}) and
(\ref{ap30}) can be rewritten in a simpler form as:
\begin{eqnarray}
\Omega_{11}(k,\tau) &=& -\Biggl( \biggl(\cos\biggl(\frac{(\sigma +
\gamma)}{2}\tau\biggr) + \frac{\Lambda^2}{(\sigma +
\gamma)}\sin\biggl(\frac{(\sigma + \gamma)}{2}\tau\biggr)\biggr)h_1
+ \nonumber\\ && + \biggl(\cos\biggl(\frac{(\sigma -
\gamma)}{2}\tau\biggr) + \frac{\Lambda^2}{(\sigma -
\gamma)}\sin\biggl(\frac{(\sigma - \gamma)}{2}\tau\biggr)\biggr)h_2
+ \nonumber\\ && +
8\,t_a\,\sin\biggl(\frac{\sigma\,\tau}{2}\biggr)\sin\biggl(\frac{\gamma\,\tau}{2}\biggr)
+ \nonumber\\ && +
4\,t_a\,\Lambda^2\biggl(g_1\sin\biggl(\frac{(\sigma +
\gamma)}{2}\tau\biggr) + g_2\sin\biggl(\frac{(\sigma -
\gamma)}{2}\tau\biggr)\biggr)\Biggr)\frac{e^{\frac{-\Lambda^2}{2}\tau}}{8\,\sigma\,\gamma},
\end{eqnarray}
\begin{eqnarray}
\Omega_{12}(k,\tau) = \Omega_{21}(k,\tau) &=&
\frac{t_b}{2\,\sigma\,\gamma}\Biggl(\frac{\Lambda^2}{(\sigma +
\gamma)}\sin\biggl(\frac{(\sigma + \gamma)}{2}\tau\biggr) -
\frac{\Lambda^2}{(\sigma - \gamma)}\sin\biggl(\frac{(\sigma
-\gamma)}{2}\tau\biggr) + \nonumber\\ && -
2\,\sin\biggl(\frac{\sigma\,\tau}{2}\biggr)\sin\biggl(\frac{\gamma\,\tau}{2}\biggr)
\Biggr)e^{\frac{-\Lambda^2}{2}\tau},
\end{eqnarray}
where
\begin{eqnarray}
h_1 &=& -(\sigma + \,\gamma)^2 - \Lambda^4,  \\
h_2 &=&  (\sigma - \,\gamma)^2 + \Lambda^4,  \\
g_1 &=& i - \frac{2\Lambda^2}{(\sigma + \,\gamma)}, \\
g_2 &=& i - \frac{2\Lambda^2}{(\sigma - \,\gamma)},
\end{eqnarray}
and, as before, $\Omega_{22}(k,\tau) = \Omega_{11}(k,\tau;
t_a\rightarrow t_d)$. Let us consider the convergence of the
stochastic process. In order for our stochastic process to converge,
the retarded Green function for the diffusion problem should obey
$G_{ab}(k, \tau)|_{\tau\rightarrow\infty}\rightarrow 0$. In other
words, we must have
$\Omega_{ab}(k,\tau)|_{\tau\rightarrow\infty}\rightarrow 0$. From
these last expressions, it is easy to see that the stochastic
process will converge, if the quantities $\sigma$ and $\gamma$ are
real, as imposed before. This lead us to the following conditions:
$|u| - y_1 > 0$ and $|u| + y_1 - \Lambda^4 > 0$, or, combining those
requirements, $|u| > \frac{\Lambda^4}{2}$. Remembering Eq.
(\ref{ap16a}), we will have the following convergence criterium:
\begin{equation}
\frac{\,\epsilon(1 + 2\sinh^2\theta)((k^2 - m^2)^2 +
\epsilon^2)}{(k^2 - m^2)^2 + \epsilon^2(\cosh^4\theta +
\sinh^4\theta) - \frac{\epsilon^2}{2}\,\sinh^{2}2\theta} >
\frac{\Lambda^2}{2},
\end{equation}
Now let us present the quantity $y_1$. As was stated before, $y_1$
is a real root of a cubic equation:
\begin{equation}
z^3 + b_2\,z^2 + b_1\,z + b_0 = 0. \label{ap37}
\end{equation}
Comparing Eqs. (\ref{ap16}), (\ref{ap17}), (\ref{ap28}) and
(\ref{ap37}), we have the following identifications: $b_2 = u -
\Lambda^4$, $b_1 = -2\,\Lambda^4\,u - \,4\,v$ and $b_0 = -4\,u\,v -
\,u^2\,\Lambda^4$. If we let:
\begin{eqnarray}
q &=& \frac{1}{3}b_1 - \frac{1}{9}b_{2}^{2},  \\
r &=& \frac{1}{6}(b_1b_2 - 3b_0) - \frac{1}{27}b_{2}^{3},
\end{eqnarray}
we will have that
\begin{eqnarray}
q &=& -\frac{4}{9}\Lambda^4\,u - \frac{\Lambda^8}{9} - \frac{4}{3}v - \frac{u^2}{9},  \\
r &=& \frac{4}{9}\Lambda^8\,u + \frac{2}{3}\Lambda^{4}\,v +
\frac{1}{18}\Lambda^4\,u^2 + \frac{4}{3}uv -
\frac{1}{27}\Lambda^{12} + \frac{1}{27}u^3.
\end{eqnarray}
So, writing $s_1 = (r + \sqrt{q^3 + r^2})^{1/2}$ and $s_2 = (r -
\sqrt{q^3 + r^2})^{1/2}$, we have that:
\begin{equation}
y_1 = (s_1 + s_2) + \frac{\Lambda^4 -\,u}{3}.
\end{equation}
As one can see, $y_1 > 0$. Also, in the limit $\epsilon\rightarrow
0$, $y_1$ becomes a polynomial of $\Lambda$.

\end{appendix}

\end{document}